\documentclass[pre,aps,floats,superscriptaddress,floatfix,twocolumn]{revtex4}
\usepackage{amssymb,amsmath}
\usepackage{graphicx}
\usepackage{psfrag}
\usepackage{color}
\usepackage{soul}
\usepackage{dcolumn}
\usepackage{bm}
\usepackage{soul}
\usepackage{cancel}
\usepackage{comment}
\usepackage[normalem]{ulem}

\def\beq{\begin{equation}}
\def\eeq{\end{equation}}
\def\bea{\begin{eqnarray}}
\def\eea{\end{eqnarray}}

\usepackage{hyperref}
\hypersetup{
  colorlinks=true,
  citecolor=blue,
  linkcolor=cyan,
}

\begin{document}
\title{ Death  by mutants: unusual multicritical dynamics   in a two-species model for absorbing state transitions}

\author{Astik Haldar}\email{astik.haldar@gmail.com}
\affiliation{Department of Theoretical Physics $\&$ Center for Biophysics,
Saarland University, 66123 Saarbr\"ucken, Germany}
\author{Abhik Basu}\email{abhik.123@gmail.com, abhik.basu@saha.ac.in}
\affiliation{Theoretical Physics Division, Saha Institute of
Nuclear Physics, 1/AF Bidhannagar, Calcutta 700064, India}

\begin{abstract}
We explore the phases and active-to-absorbing state phase transitions (AAPT) in a two-species model, where the species A and its mutant B are {\em asymmetrically} or  {\em nonreciprocally} coupled. We identify a multicritical point that connects the global absorbing state of both the species and its mutant, with a uniform active state. The critical dynamics at this multicritical point is studied within the lowest order perturbation theory. 
This asymmetric coupling between species A and mutant B leads to unequal and distinct upper critical dimensions $d_c^A$ and $d_c^B$ respectively for A and B dynamics. 
We show that the multicritical point in this model is characterised by an unusual breakdown of scale-invariance by fluctuation-induced logarithmic modulations with power law behaviour of the order parameter and correlation lengths of the species A at all dimensions  $d<d_c^A$. Above $d_c^A$, conventional scale-invariance is restored. The dynamics of the mutant species B belongs to the DP universality class with an upper critical dimension of $d_c^B=4$.

\end{abstract}

\maketitle

\section{Introduction}
The study of Phase transitions remain one of the central interest to condensed matter systems as well as the prototypical statistical mechanics models. Time-dependent correlation functions of the local order parameters at the second order or continuous phase transitions near the critical points in equilibrium systems are associated with universal dynamic scaling  parametrised by a set of universal scaling exponents that depend only on the symmetry, space dimensions and the presence or absence of conservation laws~\cite{halperin,chaikin}.
Dynamic critical phenomena in phase transitions in nonequilibrium systems that necessarily violate detailed balance are characterised by distinct nonequilibrium universality classes. Prominent examples of such nonequilibrium universality classes include    roughening transitions in the Kardar-Parisi-Zhang equation for growing nonequilibrium surfaces at dimension $d>2$~\cite{kpz,stanley,uwe} and its generalisations~\cite{diffxy-short,diffxy-long,astik-qkpz,debayan-gkpz,debayan-aniso} and the simple epidemic process with recovery or the Gribov process~\cite{gri1,gri2}. The latter, also  known as the Reggeon ﬁeld theory~\cite{reg1,reg2}, is a stochastic multiparticle process
that describes the essential features of local growth processes
of populations in a uniform environment near their extinction
threshold~\cite{ext1,ext2},  and belongs to the directed percolation (DP) universality class~\cite{hinrichsen,dp1,dp2}. In extinction transitions or active-to-absorbing state phase transitions (AAPT) belonging to the DP universality class, the population vanishes identically in the absorbing state. Since then several variants and generalisations of the simple DP model have been studied, e.g.,   systems with multiple absorbing states ~\cite{absorbing-munoz, absorbing-janssen,absorbing-Wijland}, branching and annihilation of random walkers~\cite{bran-ann-Cardy}, weak and strong dynamic scaling in a extinction transition at fluctuating fluid environment~\cite{niladri1}, role of feedback in two-species mutation~\cite{niladri-jstat,niladri2}, collective behaviour of cells with chemotactic properties in the presence of nutrient chemicals~\cite{Chemotaxis-ramin} and autochemotactic mechanism with effective long range interactions~\cite{autochemo}.

 In this article, we study the phases and the active-to-absorbing state phase transition (AAPT) in a system consisting of two interacting species $A$ and $B$, which may be interpreted as a species $(A)$ and its mutant $(B)$. This was originally proposed as a prototype model for immunization of epidemic process \cite{mutation-original}. { Due to the two species mutually interacting,  is different from the uni-directional hierarchical model~\cite{unicoupled-DP}}. We use it to study the effects of the {\em asymmetric} or {\em nonreciprocal} mutual interactions of A and B on the phases and the active-to-absorbing state transition. We show that this model admits a multicritical point, across which {\em both} species A and mutant B undergo AAPT. We use the lowest order dynamic renormalisation group (RG) to calculate the scaling exponents that characterise the AAPT.
 We find that due to the asymmetric interactions between $A$ and $B$, the AAPTs of A and B are characterised by {\em distinct}  universal scaling properties. Indeed, the upper critical dimensions of the two AAPTs are also distinct: our analysis shows that the upper critical dimension for the AAPT of $A$, $d_c^A=6$, whereas the same for the AAPT of $B$, $d_c^B=4$. This means, while fluctuations are unimportant for the AAPT of B for $d>4$, they continue to remain important for the AAPT of A till $d=d_c^A=6$.
{\ Below this $d_c^A$, the fluctuations not only corrects the critical exponents from their MF values, but {\em also} introduce {\em logarithmic modulations} to the scaling of the correlation functions, breaking conventional scale-invariance. Such logarithmic modulations are known to occur at the critical dimensions; see, e.g., Ref.~\cite{diffxy-short,log-John}.  Diluted contact processes also find the violation of the the power law due to logarithmic corrections~\cite{dcp1,dcp2}}
 
{ The rest of this article is organised as follows: In Section~\ref{model}, we discuss the discrete model defined on a lattice and its continuum representation. Before to progess the study of the phase transitions of its we discuss the popular DP universality class~\ref{DP-sec}, then we discuss the phases in the model and transitions among them in Section~\ref{phase-trans}. The behavior of the AAPT transition through multicritical point, main focus of our study, is presented in Section~\ref{MP-univ} and finally, we close the discussuon in Section~\ref{summ}. Further, some technical details are given in Appendix~\ref{actionApp} and Appendix~\ref{RGApp}.}
 
 \section{Model}\label{model}


The dynamical update rules that define the model are as follows:

(i) Self-destruction of $A$: $ A \rightarrow { \oslash}$,

(ii) Self-destruction of $B$: $ B \rightarrow { \oslash}$,

(iii) Self-production of $A$: ${ A }\rightarrow { A}+{ A}$,

(iv) Self-production of $B$: ${ B}\rightarrow { B}+{ B}$,

(v) Coagulation of $A$: ${ A}+{ A}\rightarrow { A}$,

(vi) Coagulation of $B$: ${ B}+{ B}\rightarrow { B}$,

(vii) Production of $B$ by $A$: $A\rightarrow A+B $,

(viii) Annihilation of $A$ by $B$: ${ A}+{ B}\rightarrow { B}$.

Notice that rules (i)-(vi) {\em do not} couple the dynamics of $A$ and $B$, but rules (vii) and (viii) do couple the dynamics of $A$ and $B$. Therefore, rules (vii) and (viii) should be dropped in the decoupled limit of the model. Rule (vii) signifies birth of a defective (mutant) species $B$ from the original species $A$, which in turn can consume $A$, as represented by rule (viii).

  To proceed further, we re-express the above dynamical rules as two coupled noisy Langevin equation~\cite{dp2} for the two local population densities $\rho_A$ and $\rho_B$, respectively, of the species $A$ and $B$  . The model equations are
\begin{subequations}
 \begin{align}
& \partial_t \rho_A=   D_A\nabla^2\rho_A+a_1\rho_A-a_2\rho_A^2-\mu \rho_A\rho_B +\xi_A\sqrt{\rho_A},\label{A-eqn}\\
& \partial_t\rho_B=  D_B\nabla^2\rho_B+b_1\rho_B-b_2\rho_B^2+\lambda\rho_A+\xi_B\sqrt{\rho_B}.\label{B-eqn}
\end{align}
\end{subequations}
Equations~(\ref{A-eqn}) and (\ref{B-eqn}) are generalisations of the corresponding mean-field version in Ref.~\cite{mutation-original}, { could be called as noisy coupled {\em Fisher-Kolmogorov} equations~\cite{fisher1937,Kolmogorov1988} for {\em mutation}}. { The term with coefficient $a_1$ is corresponding to the effective rate from combination of the rules (i) and (ii), with $a_2$ rate related to coagulation rule (v) in the eq.~(\ref{A-eqn}), and $b_1,\, b_2$ also similarly imply rule of combination (ii),(iv) and (vi), respectively in eq.~(\ref{B-eqn}). Further, $\lambda$ and $\mu$ terms are corresponding to the rules (vii) and (viii), respectively, which represent the corresponding feedback between two responsive species.} To allow for spatial variations and fluctuations, Eqs.~(\ref{A-eqn}) and (\ref{B-eqn}) contain diffusion of species $A$ and $B$, with diffusivities $D_A, D_B>0$ respectively, { and the noises $\xi_A$ and $\xi_B$ multiplying with the local population, which adresses no population generation at void space ensuring the population remains an absorbing state of the noisy Langevin equations~\cite{dp2}}. {Other parameters $a_2,\, b_2,\, \mu,\, \lambda$ are assumed to be positive in this model}. Here, $\xi_A$ and $\xi_B$ are assumed to be spatio-temporally white noises with zero-mean and Gaussian-distributed having variances
\begin{subequations}
 \begin{align}
 &\langle \xi_A ({\bf x},t)\,\xi_A ({\bf 0},0) \rangle= 2D_1 \delta^d({\bf x}) \delta (t),\\
 &\langle \xi_B ({\bf x},t)\,\xi_B ({\bf 0},0) \rangle= 2D_2 \delta^d({\bf x}) \delta (t).
 \end{align}
 \end{subequations}
 In the decoupled limit $\mu=0=\lambda$, both $\rho_A$ and $\rho_B$ obey similar dynamical equations, each with AAPT belonging to the DP universality class; { see next Section for detailed properties of the DP universality class}. Lastly, the model studied here is distinct from the one studied on the mutation model in Ref.~\cite{niladri2}.
  
  \section{DP Universality: Reaction Diffusion Problem}\label{DP-sec}
  Consider the dynamics of a system that consists one species $\rho$, where population density obeys only the rules ((i), (iii), (iv)) stated above. In that case, the associated reaction-diffusion process is expressed by the following Langevin equation with a multiplicative noise \cite{hinrichsen,uwe}  
  \begin{equation}
   \partial_t \rho=   D\nabla^2\rho+r\rho-u\rho^2+\xi\sqrt{\rho}. \label{dp-eq}
  \end{equation}
  The first term on the rhs gives diffusion of the species with density $\rho$, the second term on the rhs gives birth at rate $r$ and the third term gives annihilation due to over crowding at rate $u$. The multiplicative nature of the noise ensures an absorbing state admitted by this model.
Equation~(\ref{dp-eq}) shows a {\em continuous} AAPT at the mean-field critical point $r=0$ with diverging the correlation length. Near the transition point, the population density is being a critical field, the scaling anstz is defined as 
\begin{align}
\rho(r,|{\bf x}|, t)\equiv r^\beta\rho\bigg(\frac{|\bf x|}{\zeta_\perp}, \frac{t}{\zeta_\parallel} \bigg), 
\end{align}
 where $\beta$ is the order parameter exponent, and, $\zeta_\perp,\, \zeta_\parallel$ are spatial and temporal characteristic length scale, respectively. Additionally, the decay ($\alpha$), spatial length scale ($\nu_\perp$), temporal length scale ($\nu_\parallel$) exponents are correspondingly defined as $\rho\sim t^{-\alpha}$, $\zeta_\perp\sim r^{-\nu_\perp},\, \zeta_\parallel\sim r^{-\nu_\parallel}$. Further, two length scale are connected by dynamic exponent ($z$), which follows $\zeta_\parallel\sim \zeta_\perp^z$ and this immediately implies the relations $\nu_\parallel=z\nu_\perp$, $\beta=\nu_\parallel\alpha$. {An additional anomolous exponent $\eta$ is defined in the correlation function as $G(x)\sim \frac{1}{x^{d-2+\eta}}$, and this in turn implies the $\nu_\perp(d+z-2+\eta)/2=\beta$~\cite{janssen, dp2}.} Here, any three independent exponents could sufficiently express the set of others.

In the mean field theory (MFT) level, the density behaves as $\rho \sim r$ near the critical point and the time dependence of population shows $\rho\sim t^{-1}$ at $r=0$, so $\beta^\text{MF}=1,\, \alpha^\text{MF}=1$. Further, allowing the spatial fluctuation the charateristic spatial length scale varies as $\zeta_\perp\sim r^{-1/2}$ and the temporal scale from the dimensional ground behaves as $\zeta_\parallel\sim\zeta_\perp^2 $. Therefore, the exponents are identified as $z^\text{MF}=2,\, \nu_\perp^\text{MF}=1/2$, and the other remaining exponents are easily obtained by the above relations as $\nu_\parallel^\text{MF}=1,\, \eta^\text{MF}=0$. 

Nonlinear effects, neglected in the linear theory, modify the MFT values of scaling exponents above. The presence of the nonlinear terms preclude any exact calculations. However, na\"ive perturbation theory produces diverging corrections. To circumvent this problem,  perturbative dynamical Renormalisation Group (DRG) is used. This is standard and is well-documented in the literature;  see Refs~\cite{dedominicis,janssen,uwe} for technical details on the DRG  methods. Standard dimensional analysis shows the upper critical dimension $d_c=4$, and the scaling exponents depart from their mean field values in spatial dimension $d<d_c$. The fluctuation effects in $d>d_c$ are irrelevant, implying the the mean filed values for the exponents should hold. We perform one-loop DRG on a path integral representation of the governing dynamical equation~\eqref{dp-eq} with $\tilde{\epsilon}=d_c-d=4-d$ as the small parameter at the lowest (one-loop) order.  
We find
\begin{equation}
 z=2-\frac{\tilde{\epsilon}}{12},\, \nu_\perp=\bigg(2-\frac{\tilde{\epsilon}}{4}\bigg)^{-1} ,\, \beta=1-\frac{\tilde{\epsilon}}{6},\, \eta=-\frac{\tilde{\epsilon}}{12}.\label{dp-exp}
\end{equation}

  \section{Phase transitions in the two-species model}\label{phase-trans}
  
  The model equations (\ref{A-eqn}) and (\ref{B-eqn}) admit two distinct AAPTs. In the first one among these two, both the populations of $A$ and $B$ undergo absorbing state transitions, i.e., the active state consists of finite values for both $\rho_A$ and $\rho_B$, and in the absorbing state, both the densities vanish. The model has a second absorbing state in which the population of species A vanishes, but B remains in its active state.  There is, however, no absorbing state of B, in which A is in its active state. This is because the population of A serves as a ``source'' for $B$, which automatically implies that $\rho_A>0, \rho_B=0$ is not a solution of \eqref{A-eqn} and \eqref{B-eqn}. 
  The transition from a uniform active state with $\rho_A > 0,\,\rho_B>0$ to an absorbing state with $\rho_A=0=\rho_B$ takes place through a multicritical point, which in MFT is at $a_1=0=b_1$. 
  
\subsection{Mean-field theory}\label{mft}

 MFT neglects the fluctuations of the density fields. The MF equations from (\ref{A-eqn}) and (\ref{B-eqn}) follow
\begin{subequations}
 \begin{align}
  & a_1\rho_A-a_2\rho_A^2-\mu \rho_A\rho_B=0,\label{mft-A}\\
 &  b_1\rho_B-b_2 \rho_B^2+\lambda\rho_A=0. \label{mft-B}
 \end{align}
\end{subequations}

\begin{figure}[htb]
\centering
 \includegraphics[width=0.7\columnwidth]{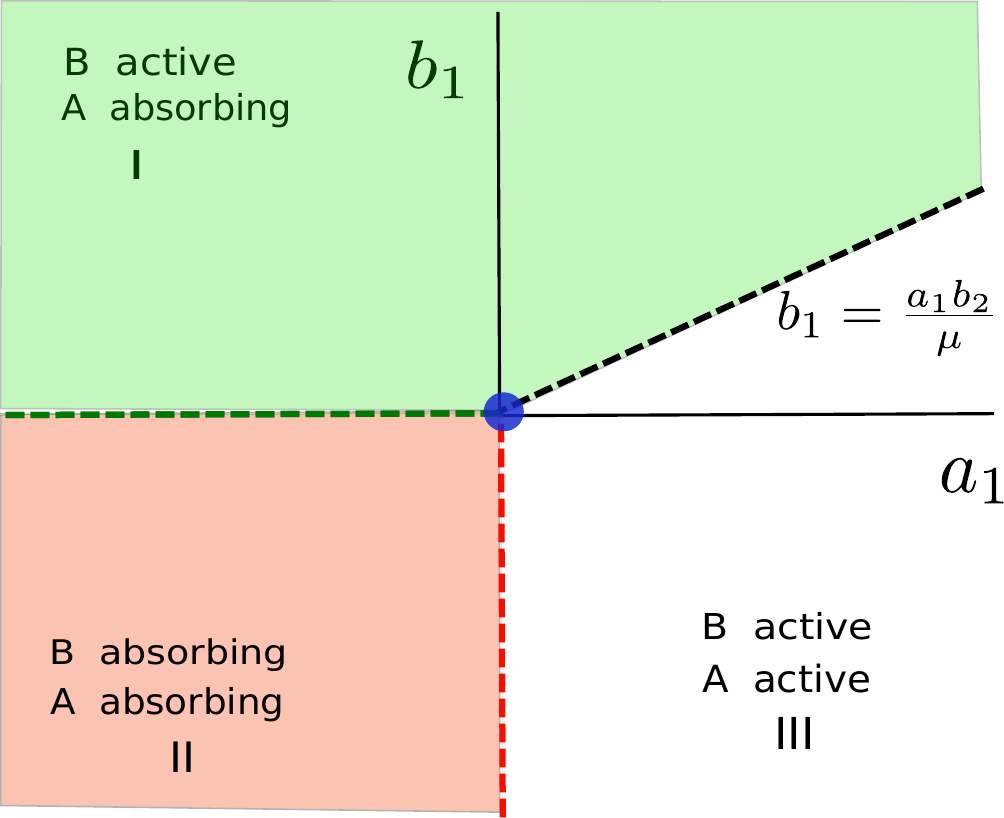}\hfill
 \caption{ Mean-field phase diagram: consists three phases as I- A absorbing B active, II- A absorbing B absorbing, III- A active-B active. Three broken lines are the trnasition lines consisting DP universality. }\label{MF-phase}
\end{figure}

The MF phase diagram is shown in Fig. \ref{MF-phase}, which has different regions marked by the parameter values, and present below the calculations of those.

{ \em Region-I; $b_1> 0, a_1<0$ :} In this parameter regime, A does not have  growing effect, so that $\rho_A^\text{MF-I}=0$  corresponding to an absorbing state of A. Further, B from Eq.~(\ref{mft-B}) is in active state with density $\rho_B^{\rm{MF-I}}=b_1/b_2>0$.

{ \em Region-II; $b_1<0 , a_1<0$ :} In this regime, neither A nor B exhibits growth implying $\rho_A^\text{MF-II}=0=\rho_B^\text{MF-II}$ from (\ref{mft-A}) and (\ref{mft-B}). Consequently, both species are in their respective absorbing states, which together constitute the global absorbing state of the model.

Given that $\rho_A^\text{MF-II}=0$ for $a_1<0$, we have $b_1=0$ is the transition line across which B undergoes DP transition since $\rho_B$ follows only DP dynamics when there are no A particles.

{ \em Region-III; $a_1>0 $:} In this case, A exhibits growth , i.e., A can be in its active state with nonzero $\rho_A^\text{MF-III}$. This in turn means B could also be in its active state due to generation effect from combination of $b_1\rho_B$ and coupling $\lambda\rho_A$. To calculate the densities in steady state, using $\rho_A=\frac{a_1-\mu\rho_B}{a_2}$ from eq. (\ref{mft-A}) into eq. (\ref{mft-B}), we obtain B density reads
\begin{align}
 \rho_B^\text{MF-III}=\bigg[(a_2b_1-\lambda\mu)\pm \sqrt{(a_2b_1-\lambda\mu)^2+4b_2\lambda a_1a_2} \bigg]/2a_2b_2.
 \end{align}
And, density of A is then $\rho_A^\text{MF-III}=\frac{a_1-\mu\rho_B^\text{MF-III}}{a_2}$. Further, the positivity of $\rho_A$ exists only the region $b_1<\frac{a_1b_2}{\mu}$, which in turn obtains $\rho_A=0$ together with $\rho_B=\frac{b_1}{b_2}$ is only acceptable solution for  $b_1\ge\frac{a_1b_2}{\mu}$ and it is same as in Region-I above. This implies Region-I spans also for $a_1>0$ with a certain values of $b_1$, and  a phase transition occurs across the line 
 \begin{equation}
 \frac{a_1b_2}{\mu}=b_1\label{A-line}
 \end{equation}
 in which A species undergoes an active to absorbing phase transition. But B is in its active state on both side of this line with density values $\rho_B^{\rm{mf-I}}$ and $\rho_B^{\rm{mf-III}}$, which coincide on the transition line. Thus, across this transition line, $\rho_B$ is {\em continuous}.  
Across (\ref{A-line}) species A undergoes a continuous transition, where A is being the critical field. In fact, since B is active on both sides of (\ref{A-line}), its fluctuations are short lived. Hence, in (\ref{A-eqn}) $\rho_B$ can be replaced by its MF active state value, giving an {\em effective} closed equation for $\rho_A$, which is indistinguishable from that for a single-species population undergoing AAPT belonging to the DP universality class.

In contrast for $ b_1<0$, B only grows due to coupled with $\rho_A$ so neglecting the higher order extinction effect $-b_2\rho_B^2$ with respect to $-b_1\rho_B$ and then B density is expressed as $\rho_B= \frac{\lambda\rho_A}{|b_1|}$ together with the value $\rho_A^\text{MF-III}=a_1/(a_2+\frac{\lambda\mu}{|b_1|})$. This in turn implies A is {\em continuously} deceases in appoarching $a_1=0,\, b_1<0$ line, so A undergoes an AAPT across the line and B also extincts on that line due to absence of A, which separates global active state to absorbing state. 
At this transition line A species is only {\em critical} allowing long lived fluctuation, whereas fluctuation of B density are short lived or {\em noncritical}, makes a closed equation for A species belonging to DP transition. Further to note, $a_1>0,\, b_1=0$ is not a transition line and the corresponding non zero densities cn be expressed by eq.~(\ref{III-densities}) using $a_1=0 $.

All the transition line, broken lines in Fig.~\ref{MF-phase}, found above meet at $a_1=0=b_1$ making a ``multicritical point'' (MP), where both species undergo active to absorbing phase transition continuously decreasing their density being both critical variabe. In the vicinity of this point $a_1\rightarrow 0^+, b_1=0$, B maintains $\rho_A=  b_2\rho_B^2/\lambda$ and $\rho_B= a_1/\mu$ with neglecting $a_2\rho_A^2$ with respect to $\rho_B$ in eq. (\ref{mft-A}) due to being low density near this transition point. 
The active state densities in the vicinity of the multicritical point satisfy
\begin{equation}
 \rho_B^{\rm{MP-III}}=\frac{a_1}{\mu},\, \rho_A^{\rm{MP-III}}=\frac{b_2\rho_B^2}{\lambda}=\frac{a_1^2b_2}{\mu^2\lambda}. \label{III-densities}
\end{equation}

 \subsection{Stability against small fluctuations of the active states}
 
 We now discuss the linear stability of the active states against their small fluctuations ignoring stochastic noise,  i.e., the densities are expressing as $\rho_A=\rho_A^{\rm{mf}}+\delta \rho_A({\bf x},t)$ and $\rho_B=\rho_B^{\rm{mf}}+\delta \rho_B({\bf x},t)$. Substituting these in Eqs. (\ref{A-eqn}) and (\ref{B-eqn}) ignoring the stochastic noises, we find the linearized equations of motion of $\delta \rho_A({\bf x},t)$ and $\delta \rho_B({\bf x},t)$ in small fluctuations limit 

\begin{equation}
  \partial_t
\begin{bmatrix}
  \delta \rho_A \\ 
  \delta \rho_B
\end{bmatrix}
= \mathcal{ M}
\begin{bmatrix}
  \delta \rho_A \\ 
  \delta \rho_B
\end{bmatrix}\label{matrix},
\end{equation}
  where the stability matrix,
  \begin{align}
  \mathcal{ M}= \begin{bmatrix}
  a_1-2a_2\rho_A^{\rm{mf}}-\mu\rho_B^{\rm{mf}}-D_A k^2 & -\mu\rho_A^{\rm{mf}} \\ 
  \lambda & b_1 -2b_2 \rho_B^{\rm{mf}} -D_B k^2 
\end{bmatrix}\nonumber
  \end{align}

Assuming the growing of the fluctuations behaves as $\delta\rho_{A,B}({\bf x},\delta t)=e^{\delta t}\delta\rho_{A,B}({\bf x},0)$ in this linear level, the eigenvalues of the linear stability matrix~(\ref{matrix}) is able to describe the growing properties of these small fluctuations~\cite{pattern-rmp, strogatz-book}. 
The negativity of the eigenvalues trivially refer the fluctuations decay about the MF densities, implying the stability of  uniform active state densities. The eigenvalues are easily given by
\begin{equation}
 \Delta_{\pm}=\frac{1}{2}[\rm{Tr}\pm\sqrt{\rm{Tr}^2-4\rm{Det}}],\label{eigenvalue}
\end{equation}
where, $\rm{Tr},\, \rm{Det}$ is the trace and determinant, respectively, of the matrix (\ref{matrix}) and are given by
\begin{subequations}
\begin{align}
 & {\rm Tr}= a_1+b_1-2a_2\rho_A^{\rm{mf}}-(\mu+2b_2)\rho_B^{\rm{mf}}- (D_A+ D_B)k^2,\label{trace}\\
 & {\rm Det}=  (D_A k^2+2a_2\rho_A^{\rm{mf}}+\mu\rho_B^{\rm{mf}}-a_1) (D_Bk^2+2b_2 \rho_B^{\rm{mf}} -b_1)\nonumber\\
 &~~~~~~~~~ +\mu\lambda\rho_A^{\rm{mf}}.\label{det}
\end{align}
\end{subequations}

One may easily identify that both eigenvalues to be negative needs $\rm{Tr}<0$ and $\rm{Det}>0$ for both $k=0$ and  $k\ne0$.

{\it In reigon-I}, B is only in active state with the densities $\rho_B=b_1/b_2,\, \rho_A=0$, the Tr and Det read
\begin{subequations}
\begin{align}
 & {\rm Tr}= a_1-b_1-\frac{\mu b_1}{b_2}- (D_A+ D_B)k^2,\label{trI}\\
 & {\rm Det}=  D_AD_Bk^4 +k^2\bigg[b_1D_A+D_B(\frac{\mu b_1}{b_2}-a_1)\bigg].\label{detI}
\end{align}
\end{subequations}
The above structure satisfies the required condition to have both negative eigenvalues for $b_1>a_1b_2/\mu$ with $a_1>0$ as well as $b_1>0$ witth $a_1<0$.

{\it For the reigon-III}, A and B both are in active state with expression \ref{III-densities}, eq. (\ref{trace}) and eq. (\ref{det}) is expressed in terms of $\rho_B^{\rm{mf}}$
\begin{subequations}
\begin{align}
 & {\rm Tr}= -a_2\rho_A^{\rm{mf}}+b_1-2b_2\rho_B^{\rm{mf}}- (D_A+ D_B)k^2,\label{trIII}\\
 & {\rm Det}=  D_AD_B k^4+k^2\bigg[D_A(2b_2\rho_B^{\rm{mf}}-b_1)+2D_Ba_2\rho_A^{\rm{mf}}) \bigg] \nonumber\\
 &~~~~~~ + (\frac{\mu\lambda}{a_2}-b_1+2b_2\rho_B^{\rm{mf}})a_2\rho_A^{\rm{mf}}.\label{detIII}
 \end{align}
\end{subequations}
Being in mind from the calculation in Sec.~\ref{mft}, $a_1-\mu\rho_B^{\rm{mf}}=2a_2\rho_A^{\rm{mf}}$ and the positive values of densities reveal $\rm{Tr}<0$ and $\rm{Det}>0$ always for $b_1<a_1b_2/\mu$ and $a_1>0$, which is the region-III itself.

 \subsection{Scaling near the multicritical point}\label{MP-univ}
 
 We now focus on the universal multicritical dynamics of the AAPT through the  multicritical point $a_1=0=b_1$. Since both the species A and B undergo continuous transitions,  their correlation lengths, $\zeta_{\perp A}, \zeta_{\perp B}$, respectively, diverge in the vicinity of the multicritical point, and hence power law scaling behaviour is expected for both of them.  Before we study this, we remind the reader that the effective equation for $\rho_A$ near the multicritical point is (\ref{A-eqn}), excluding $a_2\rho_A^2$, due to its irrelevance near the multicritical point; see Sec.~\ref{mft} and (\ref{B-eqn}) above.
  Order parameters near the multicritical point are expected to display scale invariance given by a scaling ansatz~\cite{hinrichsen}
 \begin{subequations}
 \begin{align}
  &\rho_A(a_1,|{\bf x}|,t)\equiv|a_1|^{\beta_A}\tilde\rho_A \bigg(\frac{|{\bf x}|}{\zeta_{\perp A}},\frac{t}{\zeta_{\parallel A}}\bigg),\label{A-beta-scaling}\\
  &\rho_B(b_1,|{\bf x}|,t)\equiv |b_1|^{\beta_B}\tilde\rho_B \bigg(\frac{|{\bf x}|}{\zeta_{\perp B}},\frac{t}{\zeta_{\parallel B}}\bigg).\label{B-beta-scaling}
 \end{align}
 \end{subequations}
Here, $\beta_A$ and $\beta_B$ are order parameter exponents. $\zeta_{\perp A}$,  $\zeta_{\parallel A}$ are spatial length scale and temporal length scale respectively for A species and also similarly $\zeta_{\perp B}$,  $\zeta_{\parallel B}$ for B; $\tilde\rho_a,\,\tilde\rho_B$ are dimensionless scaling functions. The characteristic length scales  diverge near the multicritical point as
\begin{align}
 & \zeta_{\perp A} \sim |a_1|^{-\nu_{\perp A}},\,  \zeta_{\parallel A} \sim |a_1|^{-\nu_{\parallel A}},\nonumber\\ 
 & \zeta_{\perp B} \sim |b_1|^{-\nu_{\perp B}}, \zeta_{\parallel B} \sim |b_1|^{-\nu_{\parallel B}},\label{length-scaleB}
\end{align}
 along with $\zeta_{\parallel A}\sim\zeta_{\perp A}^z$, which means $ \nu_{\parallel A}=z_A\nu_{\perp A}$, $z_A$ dynamic exponent for A. Similarly $\nu_{\parallel B}=z_B\nu_{\perp B}$ for B particle. Therefore, any two of the exponents are independent among three for $\rho_A$ or $\rho_B$.

In MFT, neglecting fluctuations, $\rho_B\equiv\frac{a_1}{\mu}$  and $\rho_A\sim\rho_B^2\equiv \frac{a_1^2b_2}{\mu^2}$ from Eqs. (\ref{mft-A}) and (\ref{mft-B}) { are active state densities} at vicinity of multicritical point. Therefore, { $\beta_B=1$, $\beta_A=2$}  and dynamic exponent $z_A=z_B=2$ in the linear theory allowing spatial fluctuation (i.e., when all the nonlinear terms are ignored).  
 The diffusion terms reveal $\zeta_{\perp A}\sim a_1^{-1/2}$, $\zeta_{\perp B}\sim b_1^{-1/2}$, which in turn mean the characteristic spatial length scale exponents are $\nu_{\perp A}=1/2=\nu_{\perp B}$. The decay exponents are defined by the behavior how the fields decay at MP, $\rho_A\sim t^{-\alpha_A},\, \rho_B\sim t^{-\alpha_B}$ similarly as Section~\ref{DP-sec}. 
B population decay at MP as $\rho_B\sim t^{-1}$, whereas $\rho_A $ varies differently as $\sim t^{-2}$ sincce $\rho_A\sim \rho_B^2$. Therefore, the exponents $\alpha_A=2,\, \alpha_B=1$. We thus conclude that at the mean-field level, species B dynamics belong to the MFT DP universality, whereas species A does not.

Whether or not the fluctuations neglected in MFT are ``relevant'' (in the renormalisation group or RG sense) and modify the MFT scaling behavior is a question that requires careful scrutiny.  To proceed further, it is convenient to re-express (\ref{A-eqn}) and (\ref{B-eqn}) as a path integral over configurations of $\rho_A({\bf x},t),\rho_B({\bf x},t)$ and their dynamic conjugates $\hat\rho_A({\bf x},t),\hat\rho_B({\bf x},t)$~\cite{janssen}. 
Presence of the nonlinear terms in (\ref{A-eqn})-(\ref{B-eqn}) or equivalently the anharmonic terms in the corresponding action functional preclude exact enumeration of the correlation functions, necessitating perturbative approaches. However, below the upper critical dimension, which is model specific, the anharmonic terms in the action are ``relevant'' in the RG sense, and as a result the perturbative corrections to the correlation functions can have Infrared (IR) divergence, making applications of na\"ive perturbation theories questionable. 
These perturbative corrections can however be systematically dealt within a dynamic RG framework, which is well documented in the literature~\cite{halperin,uwe}. Nevertheless, we present our RG calculations in this main text, and few more detailes of the calculations are given in Appendix~\ref{RGApp}.


  By using dimensional analysis we identify two dimensionless effective coupling constants as follows:
\begin{align}
 g_1=\frac{D_2b_2}{  D_B^2\tau_B}\varLambda^2 \tilde{k_6},\, g_2=\frac{D_1b_2\lambda}{  D_A^2  D_B\tau_A}\tilde{k_6},\, g_3=\frac{D_1\mu\lambda}{  D_A^3\tau_A}\tilde{k_6}.
\end{align}
Here $\tilde{k_6}=S_6/(2\pi)^6$, where $S_6$ is the surface area of a hypersphere of unit radius in six dimension.  Na\"ive perturbative corrections to  model parameters below the critical dimensions produce IR diverging contributions. To deal with these diverging corrections systematically, we perform one-loop Wilson momentum shell dynamic RG procedure~\cite{halperin,forster}. This involves integrating out the higher wavevectors parts of the fields and their dynamic conjugates in a path integral representation that is equivalent to the governing dynamical equations. The resulting fluctuation corrections to the different model parameters are represented by the one-loop Feynman diagrams (see Appendix~\ref{RGApp} for details). 

 Next rescale all wavevectors by ${\bf k}\rightarrow b{\bf k}$ to restore the upper cut-off and fequencies by $\omega \rightarrow b^z\omega$. The corresponding rescaling of the dynamical fields are
\begin{align}
 &\rho_A({\bf k},\omega)=b^{\chi_A} \rho_A(b{\bf k},b^z\omega),\, \hat\rho_A({\bf k},\omega)=b^{\hat\chi_A}\hat\rho_A(b{\bf k},b^z\omega) ,\\
&\rho_B({\bf k},\omega)=b^{\chi_B} \rho_B(b{\bf k},b^z\omega),\, \hat\rho_B({\bf k},\omega)=b^{\hat\chi_B}\hat\rho_B(b{\bf k},b^z\omega). \label{scaling}
\end{align}
  Under the rescaling of wavevectors, frequencies and fields, the model parameters are scaled as
 \begin{align}
  &(D_A,D_B)\rightarrow b^{z-2}(D_A,D_B),\nonumber\\  
  &(b_2,\mu)\rightarrow b^{\chi_B-d}(b_2,\mu),\, \lambda\rightarrow b^{z+\chi_A-\chi_B}\lambda,\nonumber\\
  &D_1\rightarrow b^{2z-\chi_{A}}D_1,\, D_2\rightarrow b^{2z-\chi_{B}}D_2. \label{sc-par}
 \end{align}
 
 Here, $\chi_A,\, \chi_B$ are the spatial scaling or roughness exponents the fields A and B.   Imposing the nonrescaling behavior of the model parameters, we find the MF values of the exponents which reveals $\chi_A=4=\chi_B$ from nonrescaling of $D_1,\, D_2$, and $z=2$ from nonrescaling of $D_A,\, D_B$, as before.  We thus find the dimensionless couplings scale as $g_1\rightarrow g_1b^{4-d}, g_2\rightarrow g_2b^{6-d}, g_3\rightarrow g_3b^{6-d}$ using above expressions \ref{sc-par}.
 {Hence, under rescaling both $g_2$ and $g_3$ grow with length scales  for $d<6$, giving $d_c^A=6$ as their critical dimensions, whereas $g_1$ grows under rescaling only when $d<4$, giving $d_c^B=4$ as its critical dimension. Hence, $g_1$ is {\em irrelevant} (in the RG sense) in the presence of $g_2$ and $g_3$. We further define a dimensionless ratio $\Gamma=D_B/D_A$.}

With $b>1$, $b=e^{\delta\ell}\approx 1+\delta \ell$ for small $\ell$, we obtain the following differential recursion relations:
 \begin{align}
  \frac{d\Gamma}{d\ell}=\Gamma\bigg[\frac{g_1}{d}-\frac{g_3}{d(1+\Gamma)}\bigg\{ 1+\frac{2}{1+\Gamma}+\frac{4}{(1+\Gamma)^2}\bigg\} \bigg].\label{beta-flow}
 \end{align}

 
  Since $g_1$ is {\em irrelevant} (in the RG sense) in the presence of $g_3$ as said above,  $\Gamma^*=0$ is the stable fixed point in the asymptotic long wavelength limit  according to Eq. (\ref{beta-flow}) . 
 Now, the RG recursion relations of the model parameters read
 
\begin{align}
  &\frac{d  D_A}{d\ell}=  D_A[z-2+\frac{7g_3}{d}],\, \frac{d  D_B}{d\ell}=  D_B[z-2+g_1/d],\nonumber\\ 
  & \frac{dD_1}{d\ell}=D_1[2z-\chi_A-g_3],\, \frac{dD_2}{d\ell}=D_2[2z-\chi_B-g_1],\nonumber\\
  &\frac{d\mu}{d\ell}=\mu[\chi_B-d-\frac{g_3}{2}],\, \frac{db_2}{d\ell}=b_2[\chi_B-d-3g_1/2],\nonumber\\
  &\frac{d\lambda}{d\ell}=\lambda[z+\chi_A-\chi_B-g_2].
\end{align}\label{par-flow}

 These above relations allow us to calculate the flow equations of the dimensionless coupling constants, which are given by
\begin{subequations}
 \begin{align}
 & \frac{dg_1}{d\ell}=g_1[4-d-g_1(\frac52+\frac2d)],\label{g1-flow}\\
 & \frac{dg_2}{d\ell}=g_2[6-d-g_2-g_3(1+\frac{14}{d})-g_1(\frac32-\frac1d)],\label{g2-flow}\\
 & \frac{dg_3}{d\ell}=g_3[6-d-g_2-g_3(\frac32+\frac{21}{d})].\label{g3-flow}
\end{align}
\end{subequations}
 The dynamics of species $B$ is controlled only by the coupling $g_1$ (see Appendix for the RG recursion relations of $D_B$, $D_2$). Further, the RG flow equation of $g_1$ is decoupled from other dimensionless couplings; see Eq.~(\ref{g1-flow}) above. From Eq.~(\ref{g1-flow}), $d=4$ is upper critical dimension and $g_1^*=\frac{4-d}{3}$ is the stable fixed point for $d<4$,  as in the  DP problem~\cite{hinrichsen,uwe}. Therefore, species B undergoes a continuous phase transition at the multicritical point that is characterised by a set of critical exponents belonging to the DP universality class, and associated fluctuation corrected exponents for  $d<4$  are Eq.~(\ref{dp-exp}) in Section~\ref{DP-sec}.  Clearly, $g_1$ is irrelevant at dimension $d>4$  as it flows to zero.
 
 
Furthermore for $d<4$, while $g_1^*>0$, $g_1$ is subleading in the presence of $g_2, g_3$ in the dynamics of mutant B, since $g_1$ has a $d_c^B=4$ that is less than $d_c^A=6$ for $g_2,\,g_3$. Therefore, The AAPT of species A through the multicritical point is controlled by both $g_2,\,g_3$. Thus, neglecting $g_1$ in Eq.~(\ref{g2-flow}),  to the lowest order in $\epsilon=6-d$, the flow equations of $g_2$ and $g_3$ read
 \begin{subequations}
  \begin{align}
 & \frac{d g_2}{d\ell}=g_2\bigg[\epsilon-g_2-\frac{10}{3}g_3\bigg],\label{g2-flow-new}\\
 & \frac{dg_3}{d\ell}=g_3\bigg[\epsilon-g_2-5g_3\bigg].\label{g3-flow-new}
\end{align}
\end{subequations}

\begin{figure}
 \includegraphics[width=0.7\columnwidth]{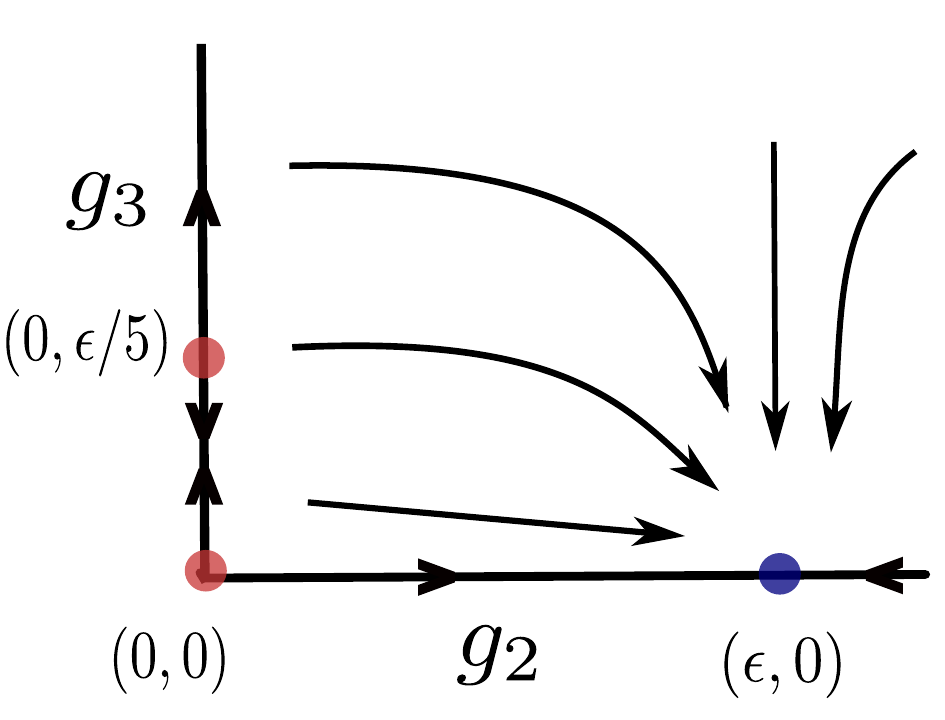}
 \caption{Schematic RG flow diagram in the $g_2-g_3$ plane ($d<d_c^A=6$). Different fixed points are marked by filled circles. Arrows indicate the direction of the RG flow (see text). }
\end{figure}

The fixed points (FPs) $(g_2^*,g_3^*)$ are found by solving Eqs.~(\ref{g2-flow-new}) and (\ref{g3-flow-new}) setting $\frac{d g_2}{d\ell}=0=\frac{d g_3}{d\ell}$, which  are $(0,0),\, (0,\frac{\epsilon}{5}),\, (\epsilon,0)$. Among the three FPs $(\epsilon,0)$ is globally stable. { While $g_3=0$ at the FP, $g_3$ approaches zero {\em slowly}, being given by}
\begin{equation}
 \frac{d g_3}{dl}=-5g_3^2.\label{flow-g_3}
\end{equation}
 This gives $g_3(\ell)= \frac{g_{30}}{1+5\ell g_{30}}$, where $g_{30}\equiv g_3(\ell=0)$ is the bare or unrenormalised value of that coupling, and  
 \begin{equation}
  g_3(\ell)\sim \frac{1}{5\ell}, \label{g3-large}
 \end{equation}
for large RG time $\ell$. 
Now using $g_{3}(\ell)$ and $z=2$ in $D_A$ flow (see Appendix), we find the renormalised or scale-dependent diffusivity 
 \begin{equation}
  D_A(\ell)\sim D_{A0} \ell^{7/30} \implies D_{A}(k) \sim D_{A0} \bigg(\ln \frac{\varLambda}{k}\bigg)^{7/30}, \label{D-k}
 \end{equation}
 where $\ell \equiv\ln ({\varLambda}/{k})$  is convenient to choose as ``RG time'' and $D_{A0}$ is bare value of $D_A$. Naturally, in the long wavelength limit $k\rightarrow 0$, $D_A(k)\gg D_{A0}$ , its value at small scales. 
Due to this logarithmic modulations,  species A is logarithmically superdiffusive.  Such fluctuation-induced logarithmic modulations of model parameters, which necessarily imply breakdown of conventional scaling, have been found in other models, but only at the {\em  critical dimensions}~\cite{diffxy-short,diffxy-long,astik-qckpz,log-John}. In contrast, here such breakdown of conventional scaling by logarithmic corrections holds at {\em all} dimensions below the upper critical 
dimension.

{ It is now expected that the correlation function of $\rho_A$ should display logarithmic modulations. In anticipation of these logarithmic corrections, we generalise the scaling ansatz (\ref{scaling}) to}
\begin{align}
 &\rho_A({\bf k},\omega)=b^{\chi_A}(\ln b)^m \rho_A(b{\bf k},b^z\omega).\label{logA-cor}
\end{align}
{ The scaling exponent 
$\hat\chi_A$ can  expressed  in terms of $\chi_A$ as defined in (\ref{logA-cor}) by imposing the condition that $\partial \rho_A/\partial t$ does not renormalise, a constraint that produces a relation between
 $\chi_A$ and $\hat \chi_A$ in Appendix~\ref{RGApp}. In the RG recursion relations for $D_A$ and $D_1$,  setting $g_3(\ell \rightarrow \infty)=0$ as the fixed point value gives $z_A=2$ and $\chi_A=4$ (equivalently $\chi_{RA}=\chi_A-d-z=-4$ using $d=6$), same as their values in the Gaussian theory. However, the ``slowness'' of the approach of $g_3(\ell)$ to zero for large $\ell$ is expected to introduce logarithmic corrections, which have their origins in the anharmonic effects, to the scaling behaviour found in the Gaussian limit.}  We further find
\begin{align}
D_1^\prime 
  & =D_1 (\ln b)^{-m} [1-g_3\ln b] \equiv D_1 b^{-g_3}(\ln b)^{-m},
\end{align}
where the above equation is obtained by using the Gaussian theory values of $z=2$ and $\chi_A=4$, and using $1-g_3\ln b \equiv b^{-g_3}$ with $b=e^{\delta l}\approx 1+\delta l$ for small $\delta l$. By exploiting the arbitrariness of the rescaling procedure, we assume $D_1$ to be unchanged under the rescaling of space, time and the fields in the same spirit of the RG procedure. Therefore, $(\ln b)^{m}=b^{-g_3}$, which gives the scaling in Eq. (\ref{logA-cor}) $b^{\chi_{A}}(\ln b)^{m}\simeq b^{4-\frac{1}{5\ln b}}$ by using the form of $g_3(\ell)$ for large $\ell$ behaviour. This gives equivalently in real space
\begin{equation}
\rho_A(|{\bf x}|,t)=b^{-4+\epsilon-\frac{1}{5\ln b}} \rho_A\bigg(\frac{|{\bf x}|}{b},\frac{t}{b^z}\bigg),\label{rho-log}
\end{equation}
giving how $\rho_A({\bf x},t)$ scales under spatio-temporal rescaling.

Let us now focus on the characteristic length scale $\xi_{\perp A}$ and examine how it scales under the rescaling of space, time and the fields.
The rescaling of $a_1$ reads ({ see Appendix~\ref{RGApp} for details}),

\begin{align}
 a_1^\prime=a_1b^z\bigg[1+\frac32 g_3 \ln b \bigg]\bigg[1-2g_3\ln b -\frac{b_1}{a_1}\ln b \bigg].\label{a1-prime}
\end{align}
Using $(b_1/a_1)^*=0$, we write the flow of $a_1$ as



\begin{align}
 \frac{da_1}{d\ell}=a_1 [z-\frac{g_{3\ell}}{2}] \implies  a_{1\ell} \simeq a_{1}(\ell=0) e^{2\ell}\ell^{-1/10}.\label{a1l}
\end{align}
The above result has been found by using {linear result} $z=2$  and the asymptotic behaviour of $g_3(\ell)$.  By using the definition in Eq. (\ref{length-scaleB}), we get
\begin{equation}
\xi_\perp^\prime \sim a_{1\ell}^{-\nu_{\perp A}}\equiv \xi_\perp(b^2 (\ln b)^{-1/10})^{-\nu_{\perp A}}, \label{xi-perp}
\end{equation}
 where the renormalized form of $a_1$ is consumed from Eq. (\ref{a1l}) and replaced by $b=e^\ell$. Now, due to this logarithmic modulations the scaling exponent cannot be extracted in the usual manner by compairing $\xi_\perp^\prime= b^{-1}\xi_\perp$ with  rescaling $x\rightarrow x/b$. 
 { Similarly the  temporal correlation also breaks power law: we find $\xi_\parallel^\prime \sim  \xi_\parallel(b^2 (\ln b)^{-1/10})^{-\nu_{\parallel A}}$ comparing with $\xi_\parallel^\prime= b^{-z}\xi_\parallel$ together with rescaling $t\rightarrow t/b^z$}.

Lastly  we work out how the order parameter varies with the distance from the multicritical point for $a_1=0^+,\, b_1=0$.
Density  $\rho_A({\bf x},t)\simeq a_{1}^{\beta_A}(\ell=0)$ as defined in Eq. (\ref{A-beta-scaling}) with the order parameter exponent $\beta_A$.  After space-time rescaling, this takes the form $\rho_A({\bf x}/b, t/b^z)\simeq a_{1\ell}^{\beta_A}$. Then the rescaling of density is by using the behaviour of $a_{1\ell}$ from Eq. (\ref{a1l}),
\begin{equation}
\rho_A(|{\bf x}|/b, t/b^z)\simeq \rho_A(|{\bf x}|, t) b^{2\beta_A} (\ln b)^{-\beta_A/10}, \label{beta-scale}
\end{equation}
which once again indicate breakdown of conventional power laws due to logarithmic modulations close to the multicritical point, valid at all dimensions below $d_c^A=6$.

\begin{table}[!htb]
\begin{center}

\begin{tabular}{|p{0.12\columnwidth}|p{0.2\columnwidth}|p{0.55\columnwidth} |}
\hline
  & MF  & RG \\
 \hline
  $\xi_{\perp A}$ & $a_1^{-1/2}$ & $a_1^{-1/2}(\varLambda/k)^{-1}(\ln(\varLambda/k))^{1/20}$\\
 \hline
 $\xi_{\parallel A}$ & $a_1^{-1}$ & $a_1^{-1}(\varLambda/k)^{-2}(\ln(\varLambda/k))^{1/10}$\\
  \hline
$\rho_A$ & $a_1^2$ & $a_1^2(\varLambda/k)^4(\ln(\varLambda/k))^{-1/5}$\\
 \hline
 $\omega(\rho_A)$ & $k^2$ & $k^2(\ln(\varLambda/k))^{-7/30}$\\
 \hline
 $\xi_{\perp B}$ & $b_1^{-1/2}$ & $b_1^{-\frac12(1+\frac{\tilde{\epsilon}}{8})}$ \\
 \hline
 $\xi_{\parallel B}$ & $b_1^{-1}$ & $b_1^{-\frac12(2+\frac{\tilde{\epsilon}}{6})}$\\
 \hline
 $\rho_B$ & $b_1$ & $b_1^{(1-\frac{\tilde{\epsilon}}{6})}$ \\
 \hline
 $\omega(\rho_B)$ & $k^2$ & $k^{(2-\frac{\tilde{\epsilon}}{12})}$\\
 \hline
 \end{tabular}
 \caption{The scaling exponents are shown MF values for dimension $d\ge 4(6)$ and RG values for dimesion  $d\le 4(6)$ for species B(A). $a_1$ in the table refers to the bare value of the parameter ($\ell=0$). $k$ represents the wavevector magnitude.}
 \label{tab1}
 
\end{center}
\end{table}

 \section{Conclusion and outlook}\label{summ}
  

 { We have studied a dynamical phase transition in a two-species model of species A and its mutant B, a prototype model of immunisation  with two absorbing phases.
We demonstrated that the transitions from active (live) phase to absorbing (dead) phases are continuous phase transitions.
When A and B are in absorbing and active state respectively, meaning the living body is recovered as healthy, showing DP universality class phase transition. The second aborbing phase with A, B both in absorbing states arises when both species undergoes AAPT through a multicritical point, shows unusual behaviour exploring the violation of power law in this continuous phase transition. 
For the second type AAPT, at the vicinity of the multicritical point B-type displays DP critical behaviour with upper critical dimension $d_c=4$ . In contrast, A shows non-DP type critical behaviour with $d_c=6$ and breaks down conventional power law formalism of continuous transition violating universality, whereas a {\em non-DP} mean-field universality exists for $d>6$. The existence of logarithmic modulation to the power laws near the critical point  in any dimension under $d_c$ for continuous phase transition has been obtained at the lowest order perturbation theory. It would be useful to go beyond the one-loop order to check the robustness of this result. 
 We expect our results will stimulate further study of continuous phase transitions of populated systems with asymmetric coupling theoretically as well as experimentally.}

\section{Acknowledgment}
A.H. thanks Alexander von Humboldt(AvH) Stiftung (Germany) postdoctoral fellowship for financial support. A.B. thanks AvH Stiftung for partial financial support through their research group linkage programme (2024) and ANRF (India) for partial financial support through the ARG (MATRICS) programme (file no.: ANRF/ARGM/2025/000461/TS).

\pagebreak

\appendix

\section{Construction of the field theory}\label{actionApp}

We first recast Eqs.~(\ref{A-eqn}) and (\ref{B-eqn}) into a path integral~\cite{dp2,janssen}. The associated dynamic generating functional is 

\begin{align}
  {\mathcal{Z}}=\int {\mathcal D}\hat\rho_A {\mathcal D}\rho_A {\mathcal D}\hat\rho_B {\mathcal D}\rho_B { e}^{-{\mathcal S_{\rm{eff}}}},\label{generating}
 \end{align}
 where  $\hat\rho_A({\bf x},t)$ and $\hat\rho_B({\bf x},t)$ are  ${\cal S}$ is the action functional
 \begin{align}
 & {\mathcal S_{\rm{eff}}}= \int_{{\bf x},t} \hat\rho_A(\tau_A\partial_t-a_1-  D_A\nabla^2)\rho_A-D_1\hat\rho_A \hat\rho_A \rho_A \nonumber\\
 &~~ +\mu\hat\rho_A\rho_A\rho_B  + \hat\rho_B(\tau_B\partial_t-b_1-  D_B\nabla^2)\rho_B \nonumber\\
 &~~ -\lambda\hat\rho_B \rho_A-D_2\hat\rho_B \hat\rho_B \rho_B.\label{action}
\end{align}
The coefficients $\tau_A$ and $\tau_B$, both of which are actually unity in Eqs.~(\ref{A-eqn}) and (\ref{B-eqn}), are introduced for the formal reasons of perturbative renormalisation group calculation (see below). Also the ratio of two is defined as $\gamma=\tau_B/\tau_A$.   Under {\em rapidity reversal} which entails time reversion $t\rightarrow -t$ together with  $\rho_A(x,t)\leftrightarrow -\hat\rho_A(x,-t), \rho_B(x,t)\leftrightarrow -\hat\rho_B(x,-t) $ Eq. (\ref{action}) is not invariant unlike the directed percolation problem~\cite{uwe}.

\begin{figure}[!htb]
 \includegraphics[width=0.8\columnwidth]{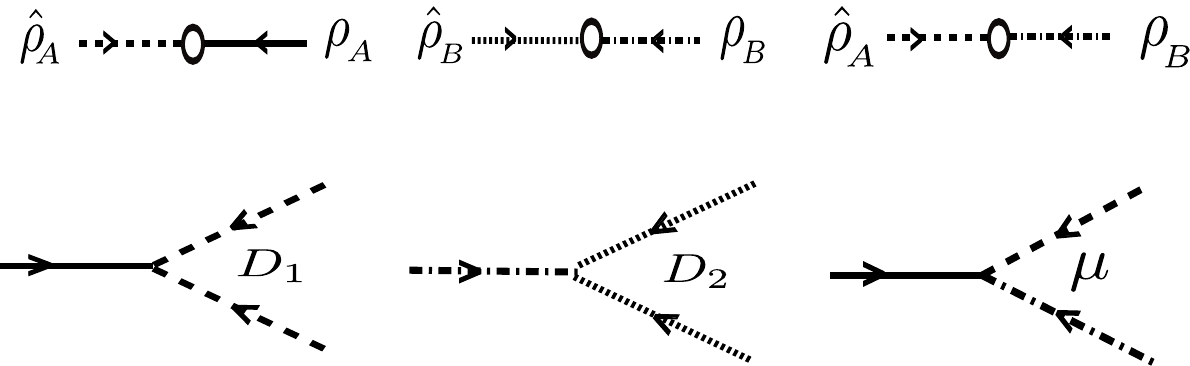}
 \caption{Top three diagrams  are the representation of the propagators in Eqs. (\ref{propagators}) and bottom three are the representative graphs of the anharmonic vertices defined in the action (\ref{action}).}\label{pro-ver}
\end{figure}

Two point functions in the Gaussian theory, calculated from the harmonic part of the action functional (\ref{action}):
\begin{subequations}
 \begin{align}
  &\langle \hat{\rho_A}_{-{\bf q},-\omega} {\rho_A}_{{\bf q},\omega} \rangle= \frac{1}{-i\omega\tau_A-a_1+  D_A q^2},\\
  &\langle \hat{\rho_B}_{-{\bf q},-\omega} {\rho_B}_{{\bf q},\omega} \rangle= \frac{1}{-i\omega\tau_B-b_1+  D_B q^2},\\
  &\langle \hat{\rho_A}_{-{\bf q},-\omega} {\rho_B}_{{\bf q},\omega} \rangle= \frac{\lambda}{(-i\omega\tau_A-a_1+  D_A q^2)(-i\omega\tau_B-b_1+  D_B q^2)}.
 \end{align}\label{propagators}
\end{subequations}
These two point functions in Eqs.~(\ref{propagators}) and the anharmonic vertices in Eq.~(\ref{action}) are graphically shown in Fig.~\ref{pro-ver}.

\section{Renormalisation Group framework}\label{RGApp}
\subsection{ Loop corrections of the model Parameters}

\begin{figure}[!htb]
 \includegraphics[width=0.8\columnwidth]{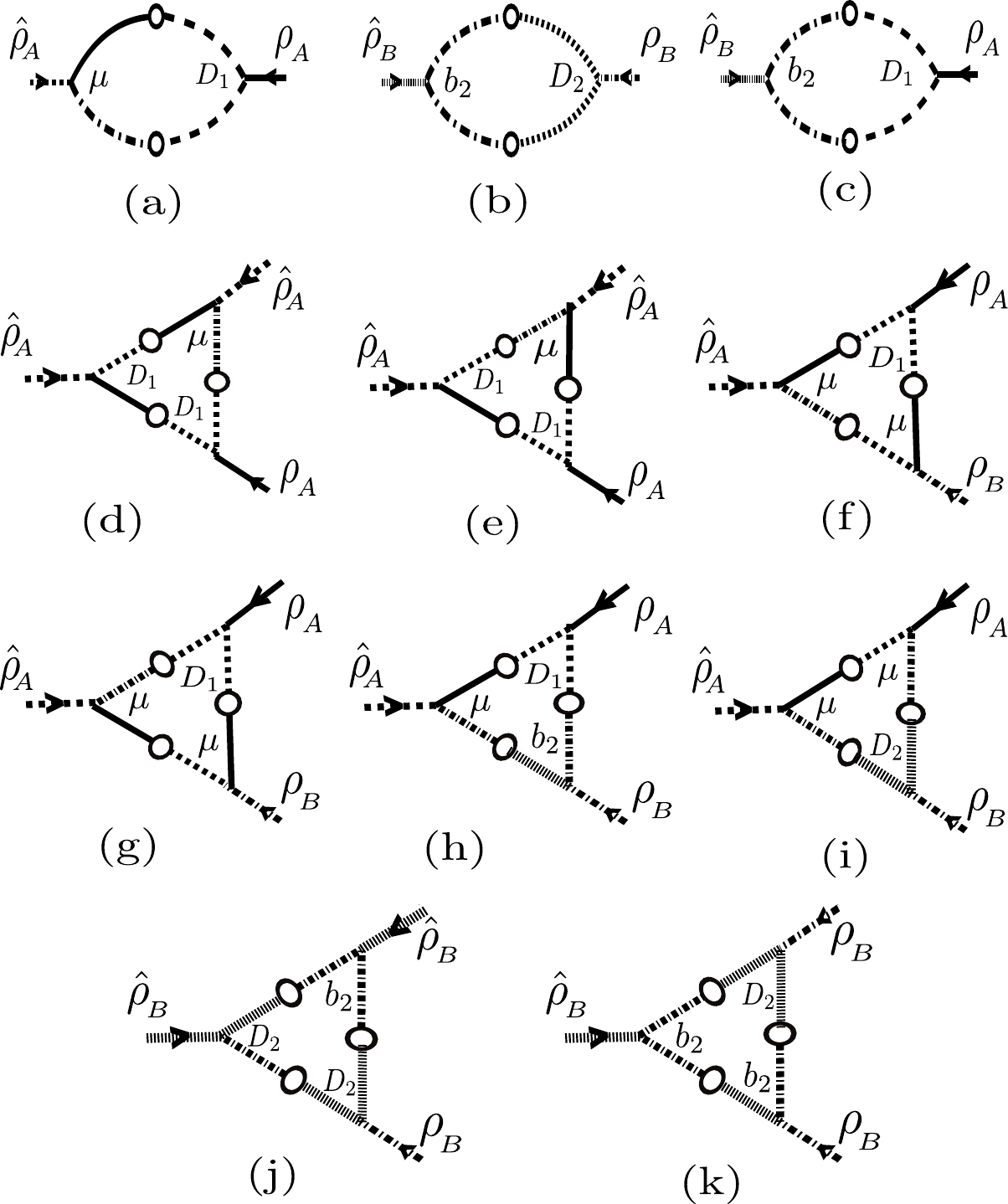}
 \caption{One-loop Feynman diagrams giving fluctuation corrections to the model parameters: (a) correction to the propagator of A (parameters: $ D_A, \tau_A, a_1$), (b) correction to the propagator of B (parametes: $ D_B, \tau_B, b_1$), (c) correction to $\lambda$, (d)-(e) corrections to $D_1$, (f)-(i) corrections to $\mu$, (j) correction to $D_2$, (k) correction to $b_2$. }\label{feynman-corr}
\end{figure}

 We treat the anharmonic terms present in Eq. (\ref{action}) in dynamic renormalisation group technique (RG), which is well documented in literature \cite{halperin,uwe}. We restrict ourselves to the lowest order perturbation theory, which is equivalent to the one-loop order. The one-loop RG calculations for the present model is very similar to the original directed percolation problem~\cite{uwe}.  For a brief,
the fields can be divided into lower momentum parts with momentum $\varLambda/b$, and higher momentum parts with momentum between $\varLambda/b$ and $\varLambda$, where $\varLambda$ is corresponding to microscopic cutoff for continuum theory and  $b>1$. Then we integrate over the momentum shell $\varLambda/b<|{\bf q}|<\varLambda$ which are representd byy diagram in Fig.~\ref{feynman-corr}.


 By evaluating the one-loop diagrams, we obtain the  RG { corrections of the model parameters}:
\begin{align}
 &\tau_A^<=\tau_A\bigg[1-\nonumber\\
 &~~~~\frac{D_1\mu\lambda}{  D_A^2(\gamma  D_A+  D_B)\tau_A}\bigg(\frac{1}{2}+\frac{\gamma  D_A}{\gamma  D_A+  D_B}\bigg)\int_{\varLambda/b}^\varLambda \frac{1}{q^6} \bigg],\label{tauA-corr}\\
 &  D_A^<=  D_A \bigg[1-\frac{D_1\mu\lambda}{  D_A^2(\gamma  D_A+  D_B)\tau_A}\bigg(\frac{1}{2}-\frac{1}{d} \nonumber\\
 &~~+\frac{\gamma  D_A}{\gamma  D_A+  D_B}(1-\frac2d) -\frac4d\bigg(\frac{\gamma  D_A}{\gamma  D_A+  D_B}\bigg)^2\bigg)\int_{\varLambda/b}^\varLambda \frac{1}{q^6} \bigg],\label{DA-corr}\\
 & a_1^<= a_1-\frac{D_1\mu\lambda}{  D_A(\gamma  D_A+  D_B)\tau_A}\int_{\varLambda/b}^\varLambda \frac{1}{q^4} \nonumber\\
 &~~ - \frac{D_1\mu\lambda}{  D_A(\gamma  D_A+  D_B)\tau_A}\bigg(\frac{a_1}{  D_A}+\frac{\gamma a_1+b_1}{\gamma  D_A+  D_B}\bigg) \int_{\varLambda/b}^\varLambda \frac{1}{q^6},\label{a1-corr}\\
 &\tau_B^<=\tau_B \bigg[1-\frac{D_2b_2}{2  D_B^2\tau_B}\int_{\varLambda/b}^\varLambda \frac{1}{q^4} \bigg],\label{tauB-corr}\\
 &  D_B^<=  D_B\bigg[1+(\frac1d-\frac12)\frac{D_2b_2}{  D_B^2\tau_B}\int_{\varLambda/b}^\varLambda \frac{1}{q^4} \bigg],\label{DB-corr}\\
 & b_1^<=b_1-\frac{D_2b_2}{  D_B\tau_B}\int_{\varLambda/b}^\varLambda \frac{1}{q^2}-\frac{D_2b_2b_1}{  D_B^2\tau_B}\int_{\varLambda/b}^\varLambda \frac{1}{q^4}\bigg],\label{b1-corr}\\
 &\lambda^<=\lambda\bigg[1-\frac{b_2D_1\lambda}{\tau_A  D_A  D_B(\gamma  D_A+  D_B)} \int_{\varLambda/b}^\varLambda \frac{1}{q^6} \bigg],\label{lam-corr}\\
 & D_1^<=D_1\bigg[1-\frac{2D_1\mu\lambda(2\gamma  D_A+  D_B)}{  D_A^2(\gamma\nu_a+  D_B)^2\tau_A} \int_{\varLambda/b}^\varLambda \frac{1}{q^6} \bigg],\label{D1-corr}\\
 &\mu^<=\mu\bigg[1- \bigg(\frac{D_1\mu\lambda(2\gamma  D_A+  D_B)}{  D_A^2(\gamma\nu_a+  D_B)^2\tau_A}\nonumber\\
 &~~ +\frac{2D_1b_2\lambda}{  D_A(\gamma\nu_a+  D_B)^2\tau_A} \bigg) \int_{\varLambda/b}^\varLambda \frac{1}{q^6}\bigg],\label{mu-corr}\\
 &D_2^<=D_2\bigg[1-\frac{2D_2b_2}{  D_B^2\tau_B}\int_{\varLambda/b}^\varLambda \frac{1}{q^4} \bigg],\label{D2-corr}\\
 &b_2^<=b_2\bigg[1- \frac{2D_2b_2}{  D_B^2\tau_B} \int_{\varLambda/b}^\varLambda \frac{1}{q^4} \bigg].\label{b2-corr}
\end{align}

We now define the corrections appeared in Eqs.~(\ref{tauA-corr}) and (\ref{tauB-corr}) for purpose of reference as
\begin{align}
 & C_{\omega A}= -\frac{D_1\mu\lambda}{  D_A^2(\gamma  D_A+  D_B)\tau_A}\bigg(\frac{1}{2}+\frac{\gamma  D_A}{\gamma  D_A+  D_B}\bigg)\int_{\varLambda/b}^\varLambda \frac{1}{q^6},\\
 & C_{\omega B}= -\frac{D_2b_2}{2  D_B^2\tau_B}\int_{\varLambda/b}^\varLambda \frac{1}{q^4}.
\end{align}

\subsection{Rescaling of the fields, Rescaled model parameters and their RG flow equations}

After integrating out the higher wavevector parts of the dynamical fields and their conjugates, the upper cut-off of the momentum is lowered to $\varLambda/b$ from $\varLambda$. By the rescaling relation relations defined in the main text Eq.~(\ref{scaling}), the upper momentum cutoff is reset to $\varLambda$.

Under the rescaling the fields, the coefficients of time derivative rescale as
\begin{align}
 &\tau_A'= \tau_A^<b^{\chi_A+\hat\chi_A}b^{-(d+2z)}\bigg[1+ C_{\omega A} \bigg],\nonumber\\
 &\tau_B'= \tau_B^<b^{\chi_B+\hat\chi_B}b^{-(d+2z)}\bigg[1+ C_{\omega B} \bigg].
\end{align}

By using the arbitrariness of the rescaling procedure, we set
 $\tau_A, \tau_B$ to remain unrenormalised~\cite{uwe} and set each of them to unity without any loss of generality. These produce the relations between $\chi_A$ and $\hat\chi_A$, and similarly between $\chi_B$ and $\hat\chi_B$. The respective relations  at the one loop level are
\begin{align}
 & b^{\chi_A+\hat\chi_A}=b^{d+2z}\bigg[1- C_{\omega A} \bigg],\, b^{\chi_B+\hat\chi_B}=b^{d+2z}\bigg[1-C_{\omega B} \bigg].\label{chiB-rel}
\end{align}

Using these above relations, we obtain the following RG recursion relations for the various model parameters:
 \begin{align}
 &   D_A^\prime=  D_A^< b^{z-2}[1-C_{\omega A}],\,~  D_B^\prime=  D_B^< b^{z-2}[1-C_{\omega B}],\nonumber\\
 & D_1^\prime= D_1^< b^{2z-\chi_A}[1-2C_{\omega A}],\,~  D_2^\prime= D_2^< b^{2z-\chi_B}[1-2C_{\omega B}],\nonumber\\
 & \lambda^\prime= \lambda^< b^{z+\chi_A-\chi_B}[1-C_{\omega B}],\nonumber\\
 & b_2^\prime = b_2^< b^{\chi_B-d}[1-C_{\omega B}],\,~  \mu^\prime= \mu^< b^{\chi_B-d}[1-C_{\omega A}],\nonumber\\
 & a_1^\prime= a_1^< b^z[1-C_{\omega A}],\,~ b_1^\prime = b_1^< b^z [1-C_{\omega B}].
\end{align}
Further, setting $b=1+\delta \ell$, differential recursion relations of $D_A$ and $D_B$ are obtained as
\begin{eqnarray}
  && \frac{d  D_A}{d\ell}=  D_A\bigg[z-2 + \frac{g_3}{d(1+\Gamma)} \bigg\{ 1+\frac{2}{1+\Gamma}+\frac{4}{(1+\Gamma)^2}\bigg\} \bigg],\label{nuA-flow1}\\
  && \frac{d  D_B}{d\ell}=  D_B\bigg[ z-2+\frac{g_1}{d} \bigg],\label{nuB-flow}
\end{eqnarray}
where $\Gamma, g_1, g_3$ are defined in the main text.
These differential equations are used to write the flow equations $\Gamma$ in the main text, and then the RG recursion relations of the model parameters.

Equivalently, the rescaling corresponds to Eq.~(\ref{scaling}) can be mapped as ${\bf x}\rightarrow {\bf x}/b,\, t\rightarrow t/b^z$ together with the fields as
\begin{align}
 &\rho_A({\bf x},t)=b^{\chi_{RA}} \rho_A({\bf x}/b,t/b^z),\,~ \hat\rho_A({\bf x},t)=b^{\hat\chi_{RA}}\hat\rho_A({\bf x}/,t/b^z) ,\nonumber\\
&\rho_B({\bf x},t)=b^{\chi_{RB}} \rho_B({\bf x}/b,t/b^z),\,~ \hat\rho_B({\bf x},t)=b^{\hat\chi_{RB}}\hat\rho_B({\bf x}/b,t/b^z). \label{scaling-real}
\end{align}
Here, $\chi_{RA}=\chi_{A}-d-z,\, \chi_{RB}=\chi_{B}-d-z$ follow from the definitions of Fourier transforms. 
We can also  re-write the flow equations of the model parameters, correspong to set of eqs.~\ref{par-flow} in main text, in terms of $\chi_{RA}$ and $\chi_{RB}$ in real space renormalisation convention, which is given by
 \begin{align}
  &\frac{d  D_A}{d\ell}=  D_A[z-2+\frac{7g_3}{d}],\, ~  \frac{d  D_B}{d\ell}=  D_B[z-2+g_1/d],\nonumber\\
  &\frac{dD_1}{d\ell}=D_1[z-d-\chi_{RA}-g_3],\,~   \frac{dD_2}{d\ell}=D_2[z-d-\chi_{RB}-g_1],\nonumber\\
  &\frac{d\mu}{d\ell}=\mu[\chi_{RB}+z-\frac{g_3}{2}],\,~  \frac{db_2}{d\ell}=b_2[\chi_{RB}+z-3g_1/2],\nonumber\\
  &\frac{d\lambda}{d\ell}=\lambda[z+\chi_{RA}-\chi_{RB}-g_2].
\end{align}\label{par-flow-R}

\bibliography{mutation.bib}

\end{document}